\documentclass[12pt]{article}
\usepackage[margin=0.95in]{geometry}

\usepackage[english]{babel}
\usepackage{amsmath}
\usepackage{lipsum}
\usepackage{ntheorem}
\usepackage{enumerate}
\usepackage{graphicx}
\usepackage{amsfonts}
\usepackage{algorithm}
\usepackage{natbib}
\usepackage{theorem}
\usepackage{appendix}
\usepackage{footnote}
\usepackage{multirow}
\usepackage{comment}
\long\def\comment#1{}  

\newtheorem*{condition*}{Condition}
\usepackage{listings}
\usepackage{color}
\usepackage{booktabs}

\begin{document}

\title{Bootstrapping F test for testing Random Effects in Linear Mixed Models}
\author{P.Y. O'Shaughnessy \and F.K.C Hui \and S M\"uller \and A.H. Welsh}
\maketitle

\begin{abstract}
Recently Hui et al. (2018) use F tests for testing a subset of random effect, demonstrating its computational simplicity and exactness when the first two moment of the random effects are specified. We extended the investigation of the F test in the following two aspects: firstly, we examined the power of the F test under non-normality of the errors. Secondly, we consider bootstrap counterparts to the F test, which offer improvement for the cases with small cluster size or for the cases with non-normal errors.
\end{abstract}

\section{Introduction}

Testing the significance of the random effects in the mixed models remains a crucial step in data analysis practice and a topic of research interest.
\citet{hui2017} revisited the F-test, which was originally proposed by \citet{wald1947} and later extended by \citet{seely1983} for testing subsets of random effects. \citet{hui2017} generalized the application to test subsets of random effects in a mixed model framework, in which correlation between random effects is allowed, and referred to this test procedure as the \emph{FLC} test. They showed that the F test is an exact test under the null hypothesis when the first two moments of the random effects are specified.

Simulations in \citet{hui2017} compared the Type I error rate and computation time for the FLC test against results of other tests for five different simulation designs. While maintaining its computational advantage, the exact FLC test consistently outperforms the parametric bootstrap likelihood ratio test and is competitive with other modern methods of random effect testing in terms of power.

Based on the promising results in \citet{hui2017}, we propose to extend the investigation of the FLC test in the following two aspects: firstly, considering that the exactness of the FLC test is conditional on the normality of the error distributions, as suggested by \citet{hui2017}, we will examine the power of the FLC test under non-normality of the errors
. Secondly, we consider a bootstrap counterpart to the FLC test, which could potentially offer improvement for the cases with small cluster size or for the cases with non-normal errors. This is inspired by the an interesting outcome shown in \citet{hui2017} that bootstrap helps with correcting power and Type I error rate for the standard likelihood ratio test when the assumption of the test is violated.

\section{Notation and Methods}

We consider examining the FLC test along with several other publicly available methods for testing random effects in linear models when the error distribution is non-normal. \citet{hui2017} considered a linear mixed model
\begin{equation}
\label{LMM}
\mathbf{y}=\mathbf{X}\boldsymbol{\beta}+\mathbf{Z}\mathbf{b}
+\boldsymbol{\epsilon}\;
\end{equation}
for an $N$-vector of observed responses $\mathbf{y}=(y_1,\ldots,y_N)$, an $N \times p$ matrix of fixed effect covariates $\mathbf{X}$ and an $N \times r$ matrix of random effects covariates $\mathbf{Z}$. The distributional assumption for the random effects is limited to the first two moments, i.e., E$(\mathbf{b})=\mathbf{0}$ and Cov$(\mathbf{b})= \boldsymbol{\Sigma}$.

To generalize \eqref{LMM} to include non-normal errors, we define $\boldsymbol{\epsilon} = \boldsymbol{\Omega}^{-1/2} \mathbf{e}$, where $\mathbf{e}$ is an $N$-vector of possibly non-normal errors which are independent and identically distrusted with mean zero and unit variance. Here $\boldsymbol{\Omega}$ is the covariance matrix for the errors. We limit the distributional assumptions on $\boldsymbol{\epsilon}$ to the first two moments, i.e., E$(\boldsymbol{\epsilon})= \mathbf{0}$ and cov$(\boldsymbol{\epsilon})=\boldsymbol{\Omega}$.

To test a subset of the random effects, we can re-write model \eqref{LMM} as
\begin{equation}
\label{LMM-part}
\mathbf{y}=\mathbf{X}\boldsymbol{\beta}+\mathbf{Z}_0 \mathbf{b}_0 + \mathbf{Z}_{-0} \mathbf{b}_{-0}
+\boldsymbol{\epsilon}\;,
\end{equation}
where $\mathbf{Z}_0$ and $\mathbf{b}_0$ denote the first $r_0$ columns and elements of $\mathbf{Z}$ and $\mathbf{b}$ for $r_0 \leq r$, and $\mathbf{Z}_{-0}$ and $\mathbf{b}_{-0}$ denote the last $(r- r_0)$ columns and elements of $\mathbf{Z}$ and $\mathbf{b}$ respectively. Suppose we wish to test
\begin{equation*}
\mbox{H}_0: \boldsymbol{\Sigma}= \left(
                            \begin{array}{cc}
                              \boldsymbol{\Sigma}_{11} & \mathbf{0} \\
                              \mathbf{0} & \mathbf{0} \\
                            \end{array}
                          \right)\;
\mbox{vs}\;
\mbox{H}_a :   \boldsymbol{\Sigma}= \left(
                            \begin{array}{cc}
                              \boldsymbol{\Sigma}_{11} & \boldsymbol{\Sigma}_{12} \\
                              \boldsymbol{\Sigma}_{12}^T & \boldsymbol{\Sigma}_{22} \\
                            \end{array}
                          \right)\;,
\end{equation*}
where $rk$ is the rank of the matrix. Here $\boldsymbol{\Sigma}_{11}$ is a $r_0 \times r_0$ non-zero and positive-definite matrix, $\boldsymbol{\Sigma}_{12}$ and $\boldsymbol{\Sigma}_{22}$ are non-zero matrices with appropriate dimensions. The FLC test statistic for test of ``no $\mathbf{b}_{-0}$ effect'' is
\begin{equation*}
F= \frac{(\mathbf{u}_0^T \mathbf{u}_0 - \mathbf{u}_1^T \mathbf{u}_1)/ \{rk(\mathbf{X},\mathbf{Z}) -  rk(\mathbf{X},\mathbf{Z_0})\} }{(\mathbf{u}_1^T \mathbf{u}_1) / \{ N - rk(\mathbf{X},\mathbf{Z})\}}\;,
\end{equation*}
where we choose $\mathbf{u}_0 = \mathbf{L}_0^T \mathbf{y}$ and $\mathbf{L}_0$ is an $N \times rk(\mathbf{L}_0)$ matrix with $rk(\mathbf{L}_0) = N- rk(\mathbf{X},\mathbf{Z_0})$, satisfying $\mathbf{L}_{0}^{T} \boldsymbol{\Omega} \mathbf{L}_0 = \mathbf{I}_{N- rk(\mathbf{X},\mathbf{Z_0})}$ and $\mathbf{L}_0 \mathbf{L}_{0}^{T}= \mathbf{I}_N - \mathbf{P}_{\mathbf{X},\mathbf{Z}_0}$. Similarly, we choose $\mathbf{u}_1 = \mathbf{L}_1^T \mathbf{y}$, where $\mathbf{L}_1$ is an $N \times rk(\mathbf{L}_1)$ matrix with $rk(\mathbf{L}_1) = N- rk(\mathbf{X},\mathbf{Z})$, satisfying $\mathbf{L}_{1}^{T} \boldsymbol{\Omega} \mathbf{L}_1 = \mathbf{I}_{N- rk(\mathbf{X},\mathbf{Z})}$ and $\mathbf{L}_1 \mathbf{L}_{1}^{T}= \mathbf{I}_N - \mathbf{P}_{\mathbf{X},\mathbf{Z}}$.

\citet{hui2017} showed that the exactness of the FLC test depends on the normality assumption of the $\epsilon$. When the normality assumption no longer holds, the FLC test statistics no longer have an F distribution in finite samples. We consider bootstrapping the FLC test. This is  inspired by the promising results shown in the parametric bootstrap standard likelihood score test in both normal and non-normal error cases.

\newsavebox{\smlmati}
\savebox{\smlmati}{$\left(\begin{smallmatrix}1&0\\0&1\end{smallmatrix}\right)$}
\newsavebox{\smlmatii}
\savebox{\smlmatii}{$\left(\begin{smallmatrix}0.05&0.02\\0.02&0.05\end{smallmatrix}\right)$}
\newsavebox{\smlmatiii}
\savebox{\smlmatiii}{$\left(\begin{smallmatrix}0.08&0.02\\0.02&0.08\end{smallmatrix}\right)$}
\newsavebox{\smlmativ}
\savebox{\smlmativ}{$\left(\begin{smallmatrix}0.10&0.05\\0.05&0.10\end{smallmatrix}\right)$}

\newsavebox{\smlmatit}
\savebox{\smlmatit}{$\left(\begin{smallmatrix}0&0\\0&0\end{smallmatrix}\right)$}
\newsavebox{\smlmatiit}
\savebox{\smlmatiit}{$\left(\begin{smallmatrix}0.2&0.1\\0.1&0.2\end{smallmatrix}\right)$}
\newsavebox{\smlmatiiit}
\savebox{\smlmatiiit}{$\left(\begin{smallmatrix}0.5&0.1\\0.1&0.5\end{smallmatrix}\right)$}
\newsavebox{\smlmativt}
\savebox{\smlmativt}{$\left(\begin{smallmatrix}1.0&0.2\\0.2&1.0\end{smallmatrix}\right)$}

\paragraph{Bootstrap Hypothesis test for FLC}
Bootstrap hypothesis testing has received considerable attention in both the theoretical and application literature.(lit review here: early work from Hinkley, young, fisher and Hall/ Hall and Wilson: guideline distinguishing setting for bootstrapping hypothesis testing from the confidence interval setting. Application in econometrics: davidson, davidson and MacKinnon )

\citet{Martin2007} summarized methods of resampling under the relevant hypotheses for some common testing scenarios. Bootstrap hypothesis testing under the linear model framework is straightforward. Firstly we define the test statistics for the observed data as $F_{obs}$. Then we construct the null resampled dataset $(\mathbf{y}^*, \mathbf{X}, \mathbf{Z})$, and obtain the bootstrap test statistics
\begin{equation*}
F^*= \frac{(\mathbf{u}_0^{*T} \mathbf{u}_0^* - \mathbf{u}_1^{*T} \mathbf{u}_1^*)/ \{rk(\mathbf{X},\mathbf{Z}) -  rk(\mathbf{X},\mathbf{Z_0})\} }{(\mathbf{u}_1^{*T} \mathbf{u}_1^*) / \{ N - rk(\mathbf{X},\mathbf{Z})\}}\;,
\end{equation*}
where $\mathbf{u}_0^* = \mathbf{L}_0^T \mathbf{y}^*$ and $\mathbf{u}_1^* = \mathbf{L}_1^T \mathbf{y}^*$. The estimated size of the bootstrap test is the proportion of bootstrap test statistics more extreme than the observed test statistics, i.e., $\# (F^* > F_{obs})/B$, where $B$ is the number of bootstrap simulations.

When constructing the bootstrap samples under the null hypothesis in this paper, we use the residual bootstrap method following \citet{PaparoditisPolitis2005}:
\begin{enumerate}
  \item  Estimate the residuals by $\hat{\boldsymbol{\epsilon}}= \mathbf{y} - \mathbf{X} \hat{\boldsymbol{\beta}} + \mathbf{Z}_0 \hat{\boldsymbol{b}_0}$;
  \item Generate the null resampled responses $\mathbf{y}^*$ by $\mathbf{y}^*= \mathbf{X} \hat{\boldsymbol{\beta}} + \mathbf{Z}_0 \hat{\boldsymbol{b}_0}+ \boldsymbol{\epsilon}^*$, where $\boldsymbol{\epsilon}^*$ is an $N$-vector with elements resampled with replacement from $\hat{\boldsymbol{\epsilon}}$.
\end{enumerate}

We also considered two other possibilities when constructing the bootstrap samples under the null hypothesis:

   \emph{Do \emph{not} impose the null hypothesis on residuals \citep{Martin2007}:} estimate the residuals by $\hat{\boldsymbol{\epsilon}}= \mathbf{y} - \mathbf{X} \hat{\boldsymbol{\beta}} + \mathbf{Z} \hat{\boldsymbol{b}}$; then generate the null resampled responses $\mathbf{y}^*$ by $\mathbf{y}^*= \mathbf{X} \hat{\boldsymbol{\beta}} + \mathbf{Z}_0 \hat{\boldsymbol{b}_0}+ \boldsymbol{\epsilon}^*$, where $\boldsymbol{\epsilon}^*$ is an $N$-vector with elements resampled with replacement from $\hat{\boldsymbol{\epsilon}}$. The difference between this bootstrap and the previous one is how fitted values and residuals are constructed. In this bootstrap method, the fitted values and residuals are constructed without setting the variance components of the testing random effects to zero.

    \emph{M-out-of-n residual bootstrap}: scale the residual by $\sqrt{n/m}$, where $n$ is the sample size and $m$ is an arbitrary number generally smaller than $n$ \citep{Shao1996}. One way to choose $m$ is to set the scale of the bootstrap residual equal to the variance ratio between the bootstrap residual and the population (true) residual. We observed that there is difference in results between the two residual bootstraps (imposing uull and not imposing null) using residuals constructed under the null and residual constructed under the alternative. A good starting point is choose $m$ satisfying $\sigma^2_{alt}/\sigma^2_0 = m/n$, in which $m$ is a ratio of variance for bootstrap with residuals not imposing null ($\sigma^2_{alt}$) and variance for bootstrap with residuals imposing null ($\sigma^2_0$). 
    Limited simulations were conducted for setting $2$ with Student's t distribution errors. The results showed large Monte-Carlo variations, and setting $m$ equal to the variance ratio does not perform better than others.


\paragraph{Double bootstrap}

Bootstrapping a quantity which has already been bootstrapped will lead to further asymptotic refinement in addition to the refinement from simply bootstrapping a test statistic once. This is the basic idea motivating the early development of the \emph{double bootstrap} in the $1980$s by several authors including \citet{Efron1983}, \citet{hall1986}, \citet{Beran1987,Beran1988}, \citet{HallMartin1988}, \citet{DiciccioRomano1988}, \citet{HinkleyShi1989}, etc.. Double bootstrap hypothesis testing is simply treating the single bootstrap P value as a test statistics and bootstrapping it again. The second-level bootstrap samples are used to construct an empirical distribution for the P value. The detailed algorithm for double bootstrapping the FLC test is given in Appendix \ref{appdx:doubleBT}. The double bootstrap P value is the proportion of the second-level P values smaller (more extreme) than the first-level P value.
This procedure is straightforward to implement, but it also comes with a high computational cost. For example, let $B_1$ and $B_2$ be the first and the second level bootstrap sample sizes. If $B_1=999$ and $B_2=499$, we need to calculate $1+B_1+B_1 B_2= 1+999+999 \times 499= 499,501$ test statistics.

One of the advantages of the original FLC test is its fast computation. The traditional double bootstrap algorithm certainly cannot retain this feature. \citet{DavidsonMacKinnon2007} proposed a less computationally demanding procedure for asymptotically pivotal test statistics. This procedure corrects P values by producing one second-level bootstrap sample for each first-level bootstrap sample to calculate a critical value at the nominal level equal to the first-level bootstrap P value. This procedure was referred as the \emph{fast double bootstrap} and the algorithm of bootstrapping the FLC test using the fast double bootstrap is as follows:
\begin{enumerate}
  \item Obtain the test statistic $F_{obs}$.
  \item Generate $B$ bootstrap samples under the null hypothesis, and compute a bootstrap statistic $F^*_k$ for each $k=1,\ldots,B$ bootstrap sample.
  \item Calculate the first-level bootstrap P value as $\hat{p}^*=\frac{1}{B} \sum_{k=1}^{B} I(F^*_k > F_{obs})$.
  \item For each of the $B$ first-level bootstrap samples, generate a single second-level bootstrap sample. Use the second-level bootstrap sample to compute a second-level bootstrap test statistic $F^{**}_{k}$.
  \item Calculate the $1- \hat{p}^*$ quantile of the $F^{**}_k$ , $\hat{Q}^{**}_{B}$, defined implicitly by the equation
      \begin{equation*}
      \frac{1}{B} \sum^{B}_{k=1} I \{F^{**}_{k}<\hat{Q}^{**}_{B} \} = 1- \hat{p}^*\,,
        \end{equation*}
        where $\hat{p}^*$ is the first-level P value defined in step 3.
  \item Calculate the \emph{fast double bootstrap} P value as
  \begin{equation*}
  \hat{p}^{**}_F= \frac{1}{B} \sum^{B}_{k=1} I \{ F^{*}_{k}> \hat{Q}^{**}_{B} \}\,.
    \end{equation*}
\end{enumerate}
The fast double bootstrap P value is the proportion of bootstrap test statistics more extreme than $\hat{Q}_{B}^{**}$, the $1-\hat{p}^*$ quantile of the second-level test statistic $F^{**}_{k}$. The number of test statistics we need to calculate for the fast double bootstrap is $1+2B$. A description of the full double bootstrap FLC test and empirical differences between the double bootstrap and fast double bootstrap are given in Appendix \ref{appdx:doubleBT}.

\section{Results}

We examine various testing methods for testing random effects in linear models when allowing correlation between random effects and not restricting the error distribution to be normal. We adopted three simulation designs from \citet{hui2017}; the details of each design along with the null hypotheses are given in Table \ref{tab:setting}. Specifically for these simulation designs, the responses $\mathbf{y}$ and the covariates $\mathbf{X}$ can be split into $i=1,\ldots,n$ clusters, such that $\sum_{i=1}^{n} m_i = N$, where $m_i$ is the cluster size for the $i$th cluster and $N$ is the total number of observations. The data are simulated from a linear mixed model
\begin{equation}
\mathbf{y}=\mathbf{X}\boldsymbol{\beta}+\mathbf{Z}\mathbf{b}
+\boldsymbol{\epsilon}\;.
\end{equation}
We define $\boldsymbol{\epsilon} = \boldsymbol{\Omega}^{-1/2} \mathbf{e}$, where $\mathbf{e}$ is an $N$-vector of independent and identically distrusted errors with mean zero and unit variance. In the simulations, we assume independent error, i.e., $\boldsymbol{\Omega}$ is a scaled identity matrix $\sigma^2 \mathbf I_N$. We limit the distributional assumption on $\boldsymbol{\epsilon}$ to the first two moments, i.e., E$(\boldsymbol{\epsilon})= \mathbf{0}$ and cov$(\boldsymbol{\epsilon})=\boldsymbol{\Omega}$. The non-normality features we consider here include spread, skewness and asymmetry. In particular, we consider the following specific cases of non-normal errors:
\begin{enumerate}
  \item Student's t distribution with 3 degrees of freedom (student);
  \item zero-mean chi-squared distribution with 3 degrees of freedom (chisq); and
  \item two-component normal mixture distribution to model contamination, with a $20\%$ contamination probability and contamination variance 9 times the core variance (2CMM).
\end{enumerate}



We set $\mathbf{Z}$ to be a block diagonal matrix with $n$ blocks. The covariance matrices for the random effects and error are also block diagonal matrices, i.e., $\boldsymbol{\Sigma}= \mathbf{I}_n \otimes \mathbf{D}$ and $\boldsymbol{\Omega}= \mathbf{I}_n \otimes \mathbf{E}$, where $\mathbf{I}_n$ denotes an $n \times n$ identity matrix and $\otimes$ denotes the Kronecker product. The random effects can be written as $\mathbf{b}= (\mathbf{b}_1^T, \ldots, \mathbf{b}_n^T)^T$, with $\mathbf{b}_i \sim N(0,\mathbf{D})$. In each simulation design, the random effects for each cluster $\mathbf{b}_i$ was generated from a multivariate normal distribution with mean zero and a defined covariance matrix $\mathbf{D}$. The simulation designs allow testing subsets of the random effects. In Setting 1, we selected $\mathbf{D}$ as one of the following four covariance matrices, ~\usebox{\smlmati}, ~\usebox{\smlmatii}, ~\usebox{\smlmatiii} and ~\usebox{\smlmativ}. In Setting 2 where we tested for the random effects 2 and 3, given the first random effect is included in the model and is independent of other two, the covariance matrix $\mathbf{D}$ was a $3 \times 3$ matrix with $D_{11}=1$, $D_{12}=D_{13}=D_{21}=D_{31}=0$ and the bottom right $2 \times 2$ submatrix $[\mathbf{D}]_2$ equaled one of the following: ~\usebox{\smlmatit}, ~\usebox{\smlmatiit}, ~\usebox{\smlmatiiit} and ~\usebox{\smlmativt}. In Setting 3, we constructed the covariance matrix $\mathbf{D}$ as a $4 \times 4$ matrix with a top left $2 \times 2$ submatrix simulated from a Wishart distribution with three degrees of freedom and a diagonal scale matrix with both elements equaled to 0.5, a bottom right $2 \times 2$ submatrix constructed to have $\tau$ as the diagonal elements and $\tau /2$ as the off diagonal elements and the remaining elements of $\mathbf{D}$ equaled $\tau /2$ to ensure the positive definiteness of $\mathbf{D}$, $\tau$ ranging over $0$, $0.1$ and $0.2$.

Based on $1000$ simulated datasets, the empirical Type I error and power were calculated using the percentage of datasets which rejected the null hypothesis in question at a nominal $5\%$ significance level. Five methods were examined: 1) the FLC test; 2) the parametric bootstrap likelihood ratio test with $999$ bootstrap samples, using the \texttt{R} package \texttt{pbkrtest} \citep{Halekoh2014}; and 3) the linear score test of \citet{Qu2013}, using the \texttt{R} package \texttt{varComp} \citep{Qu2015}; 4) the fast double bootstrap FLC test with $999$ bootstrap samples for the first bootstrap level; and 5) the residual bootstrap FLC test with $999$ bootstrap samples.

\begin{table}[h!]
\caption{Features of various simulation designs considered in this article. We use the notation $R$ to denote random effect e.g., $R1=R2=0$ means the null hypothesis tests random effects 1 and 2 are both equal to zero, and $R2=R3=0 | R1$ means the null hypothesis tests random effects 2 and 3 are equal to zero given random effect 1 is in the model. In the independent cluster designs, $n$ and $m$ refer to the number of clusters and cluster size respectively.} \label{tab:simsummary} \medskip \centering
\scalebox{0.85}{
\begin{tabular}{clcc}
\toprule[1.5pt]
Setting & Simulation Design & Sample sizes & $H_0$ \\
\midrule
1$^A$ & Independent cluster & $n = \{10,15\}; \; m = \{3,5\}$ & $R1=R2=0$\\
  & \multicolumn{3}{l}{with two fixed covariates and two uncorrelated random covariates} \\\\
2$^A$ & Independent cluster & $n = \{7,15,25,50\}; \; m = 10$ & $R2=R3=0 | R1$\\
  & \multicolumn{3}{l}{with two fixed covariates and three uncorrelated random covariates} \\\\
3$^{A,B}$ & Independent cluster & $n = \{10,20,40\}; \; m = \{10,20\}$ & $R3=R4=0 | R1,R2$\\
  & \multicolumn{3}{l}{with eight fixed covariates and four correlated random covariates} \\\\
\bottomrule[1.5pt]
\multicolumn{4}{l}{\small $^A$Simulations also performed where the random effects were not normally distributed} \\
\multicolumn{4}{l}{\small $^B$Simulations also performed testing $H_0: R4=0 | R1,R2,R3$ to include the restricted likelihood ratio} \\
\multicolumn{4}{l}{\small \hspace{5pt} test for comparison}
\end{tabular}
}
\label{tab:setting}
\end{table}

\begin{figure}[H]
 \begin{center}	
\includegraphics[width=15cm]{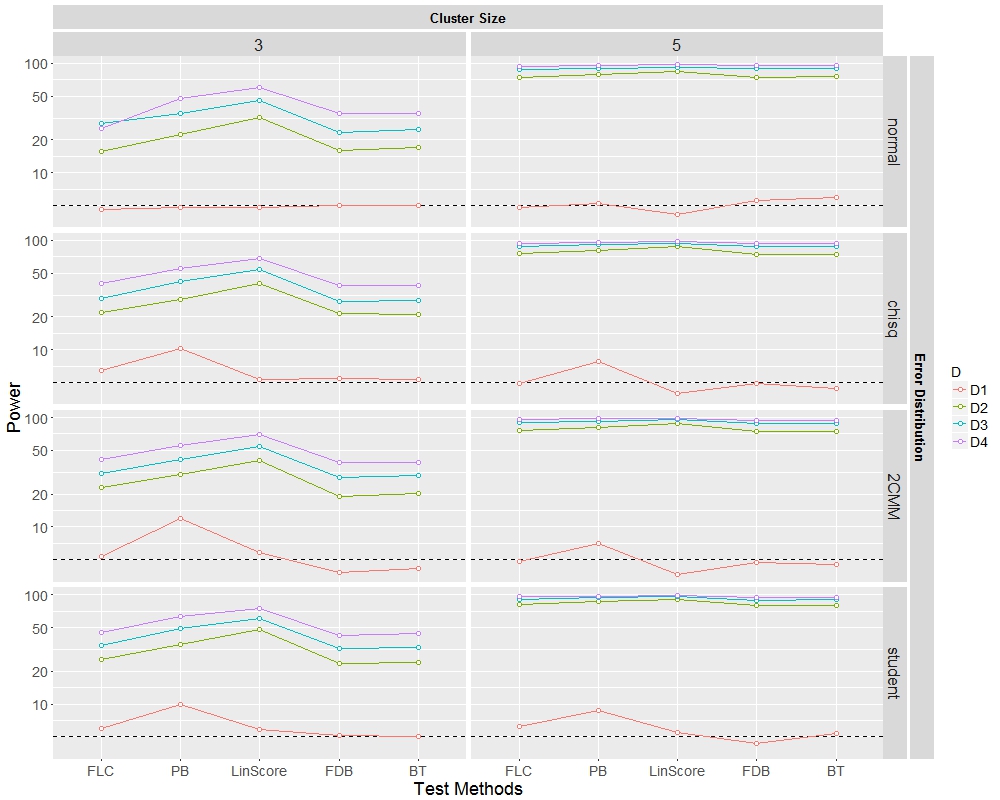}
 \end{center}
\caption{Type I error and power for Setting $1$ for four underlying distributions with different cluster sizes when the number of cluster $n=10$. The four distributions considered in the simulations are the standard normal distribution (normal), Student's t distribution with 3 df (student), the zero-mean chi-squared distribution with 3 df (chisq) and the two-component normal mixture distribution (2CMM). The methods compared included the FLC test, the parametric bootstrap likelihood ratio test (PB), the linear score (LinScore), the fast double bootstrap FLC test (FDB) and the residual bootstrap FLC test (BT). Performance was assessed in terms of percentage of datasets where the method rejected the null hypothesis. Each colored line represents one of four covariance matrices $\mathbf{D}_{1}=$~\usebox{\smlmati}, $\mathbf{D}_{2}=$~\usebox{\smlmatii}, $\mathbf{D}_{3}=$~\usebox{\smlmatiii} and $\mathbf{D}_{4}=$~\usebox{\smlmativ}. The y-axis is plotted on a log-10 scale and the dashed line is the reference line for the nominated $5\%$ significance level.}
\label{fig:s1}
\end{figure}

\begin{figure}[H]
 \begin{center}	
\includegraphics[width=15cm]{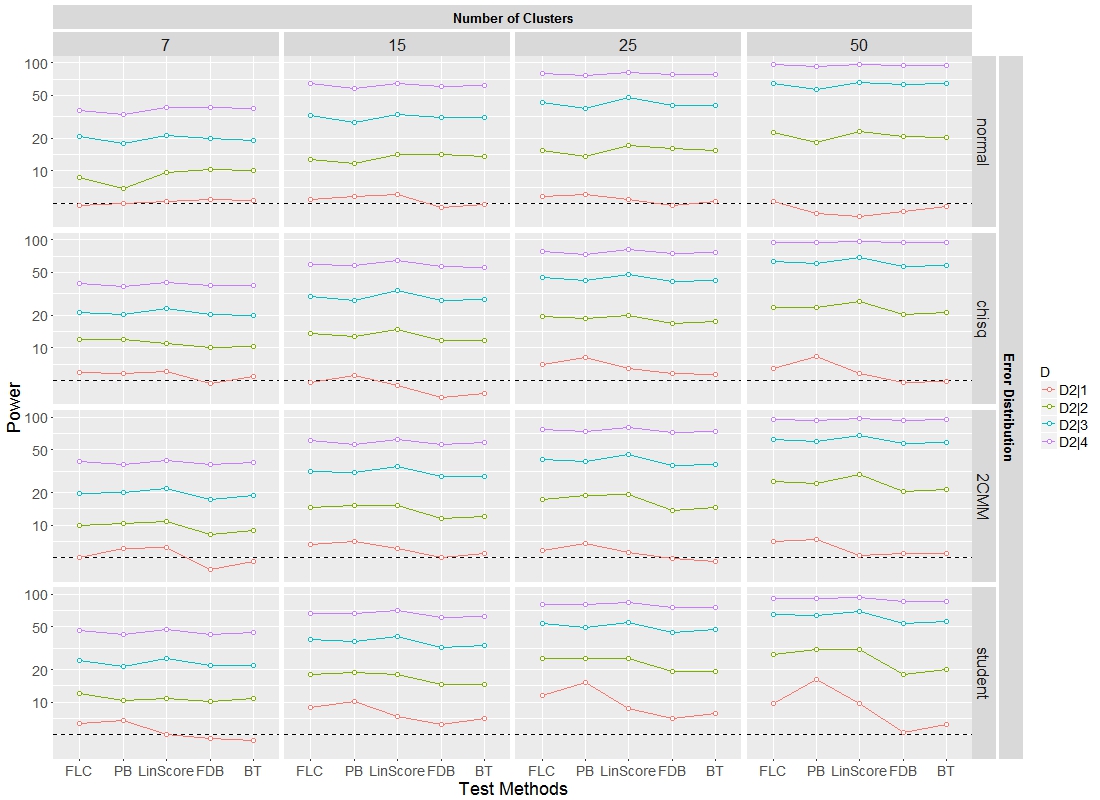}
 \end{center}
\caption{Type I error and power for Setting $2$ for four underlying distributions with different number of clusters when the cluster size $m=10$. The four distributions considered in the simulations are the standard normal distribution (normal), Student's t distribution with 3 df (student), the zero-mean chi-squared distribution with 3 df (chisq) and the two-component normal mixture distribution (2CMM). The methods compared included the FLC test, the parametric bootstrap likelihood ratio test (PB), the linear score (LinScore), the fast double bootstrap FLC test (FDB) and the residual bootstrap FLC test (BT). Performance was assessed in terms of percentage of datasets where the method rejected the null hypothesis. Each colored line represents one of four covariance matrices $\mathbf{D}_{2|1}=$~\usebox{\smlmatit}, $\mathbf{D}_{2|2}=$~\usebox{\smlmatiit}, $\mathbf{D}_{2|3}=$~\usebox{\smlmatiiit} and $\mathbf{D}_{2|4}=$~\usebox{\smlmativt}. The y-axis is plotted on a log-10 scale and the dashed line is the reference line for the nominated $5\%$ significance level.}
\label{fig:s2}
\end{figure}

\begin{figure}[H]
 \begin{center}	
\includegraphics[width=13.5cm]{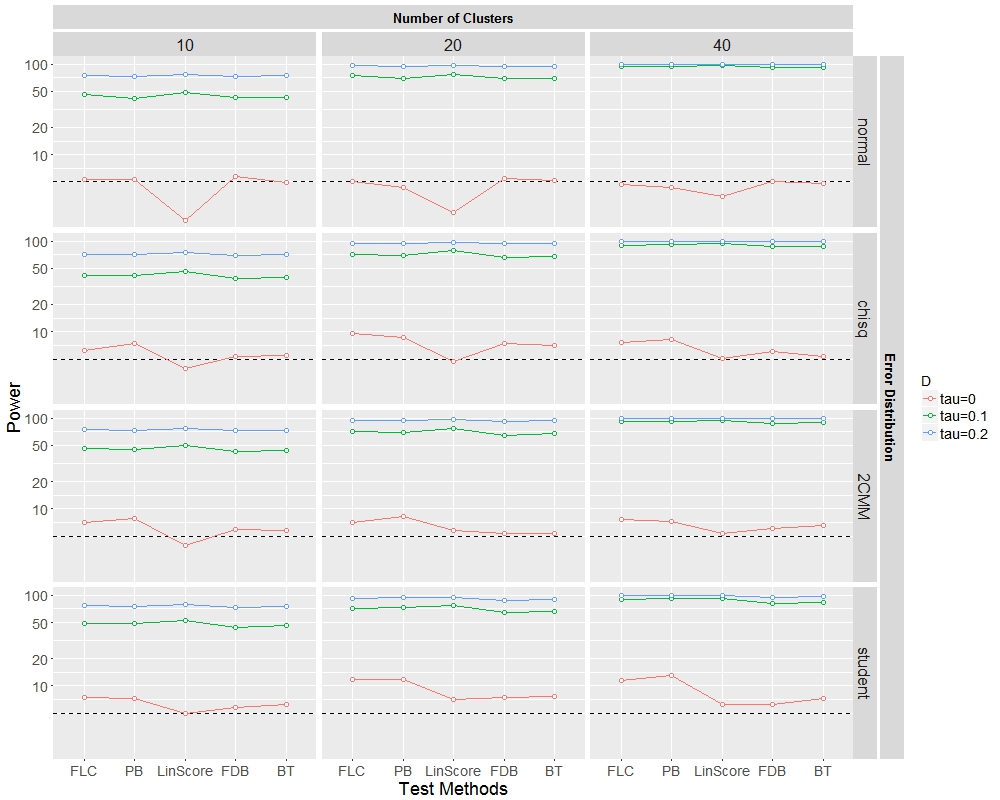}
 \end{center}
\caption{Type I error and power for Setting $3$ for four underlying distributions with different number of clusters when the cluster size $m=10$. The four distributions considered in the simulations are the standard normal distribution (normal), Student's t distribution with 3 df (student), the zero-mean chi-squared distribution with 3 df (chisq) and the two-component normal mixture distribution (2CMM). The methods compared included the FLC test, the parametric bootstrap likelihood ratio test (PB), the linear score (LinScore), the fast double bootstrap FLC test (FDB) and the residual bootstrap FLC test (BT). Performance was assessed in terms of percentage of datasets where the method rejected the null hypothesis. Each colored line represents the covariance matrices $\mathbf{D}$ defined by one of the three $\tau$. The y-axis is plotted on a log-10 scale and the dashed line is the reference line for the nominated $5\%$ significance level.}
\label{fig:s3}
\end{figure}

We compared the power results for normal error cases cited in \citet{hui2017} with our results for the non-normal error cases. Figures \ref{fig:s1}-\ref{fig:s3} contain the empirical Type I error and power for the three simulation designs shown in Table \ref{tab:setting}. Each figure compares empirical type I errors and powers for different scenarios using five methods with nominal $5\%$ significance level. A good method has the correct type I error rate and does not suffer from loss in power. Detailed results for the three non-normal error cases, along with the recorded computation time, are included in Table \ref{tab:s1-student}-\ref{tab:s3-contam} in Appendix \ref{appdx:table}. Tables \ref{tab:s1-student}-\ref{tab:s1-contam} present the results from the simulation design for setting 1, Tables \ref{tab:s2-student}-\ref{tab:s2-contam} for setting 2 and Tables \ref{tab:s2-student}-\ref{tab:s2-contam} for setting 3.

More detailed analysis of the figures leads to the following conclusions for the two proposed bootstrap FLC tests:
\begin{itemize}
  \item Two bootstrap counterparts to the FLC test, the fast double bootstrap (FDB) and the residual bootstrap (BT), outperformed the other methods when the errors are not normal in terms of empirical Type I error;
  \item These two bootstrap methods produced competitive results in term of power to other methods;
  \item The performance of the two bootstrap methods was consistent across various simulation settings.
\end{itemize}

Similar to \citet{hui2017}'s results, the linear score test still had good power in most of the settings. However its performance on Type I error rate for non-normal error cases is rather volatile. On the other hand, both residual bootstrap and fast double bootstrap FLC tests are comparable to the linear score test in term of power. Also both bootstrap FLC tests showed near-perfect results for Type I error consistently for all non-normal error cases. The fast double bootstrap FLC test is slightly better than the residual bootstrap FLC test, however it is also twice as computationally expensive as the residual bootstrap.

\comment{
\section{Asymptotic Distribution of FLC Statistic}
\begin{equation}
F= \frac{(\mathbf{u}_0^T \mathbf{u}_0 - \mathbf{u}_1^T \mathbf{u}_1)/\{ \mbox{rk}(\mathbf{X},\mathbf{Z}) -\mbox{rk}(\mathbf{X},\mathbf{Z}_0) \}}{(\mathbf{u}_1^T \mathbf{u}_1)/ \{ N -\mbox{rk}(\mathbf{X},\mathbf{Z})\}}
\end{equation}

idea: We show that the dominator $\mathbf{u}_1^T \mathbf{u}_1 / \{ N -\mbox{rk}(\mathbf{X},\mathbf{Z}) \} \overset{p}{\to} 1$ and the numerator $(\mathbf{u}_0^T \mathbf{u}_0 - \mathbf{u}_1^T \mathbf{u}_1)/\{ \mbox{rk}(\mathbf{X},\mathbf{Z}) -\mbox{rk}(\mathbf{X},\mathbf{Z}_0) \}$ has the asymptotic distribution \textcolor{blue}{normal, maybe}. Therefore the statistic $F$ has the same asymptotic distribution as the numerator completing proof of the result. we use the notation $\cdot|\mathbf{b}$ denote the distribution given the random effect coefficients $\mathbf{b}$.

$\mathbf{u}_1^T \mathbf{u}_1 / \{ N -\mbox{rk}(\mathbf{X},\mathbf{Z}) \} \overset{p}{\to} 1$
}

\comment{

\section{Literature Review Notes}
\subsection{Review on Methods}
\citet{seely1983} extended the work by \citet{wald1947} to use the Wald test to test a subsets of random effects in a linear mixed model. They gave necessary and sufficient conditions for the test to be applicable and further proposed an uniqueness property which determines whether a proposed variance component test in a mixed linear model is the Wald test.

}

\appendix
\appendixpage

\section{Simulation Results for Non-normal Error Cases}
\label{appdx:table}

\begin{table}[H]
  \centering
  \caption{Simulation results for Setting $1$ (Table 2 in \citep{hui2017}) for cases with non-normal error with Student's t distribution with 3 degrees of freedom. Methods compared included the FLC test, the standard likelihood ratio test (LRT) compared to a Chi-squared test, its parametric bootstrap (PB) counterpart, the linear score test (LinScore), bootstrap FLC test using fast double bootstrap (FDB) and residual bootstrap FLC test (BT). $D$ is the variance-covariance matrix for variance components. $m$ and $n$ are the number of clusters and cluster size, respectively. Performance was assessed in terms of percentage of datasets where the method rejected the null hypothesis, and the mean computation time in seconds(in parentheses).}
  \resizebox{\textwidth}{!}{%
\begin{tabular}{lllllllll}
    \toprule
    \textbf{D}   & n   & m   & FLC & LRT & PB  & LinScore & FDB & BT \\
    \midrule
    \multirow{4}{*}{$\left(
                      \begin{array}{cc}
                        0 & 0 \\
                        0 & 0 \\
                      \end{array}
                    \right)$}
       & 10  & 3   & 6 (0.02) & 5.7 (0.08) & 10 (38.54) & 5.9 (0.05) & 5.2 (7.82) & 5.1 (3.89) \\
       & 10  & 5   & 6.3 $(<0.01)$ & 3.1 (0.04) & 8.8 (39.06) & 5.5 (0.03) & 4.4 (7.92) & 5.4 (4.05) \\
       & 15  & 3   & 8 $(<0.01)$ & 11.4 (0.04) & 15.8 (39.87) & 6.8 (0.03) & 4.8 (8.03) & 5.2 (3.99) \\
       & 15  & 5   & 9.7 $(<0.01)$ & 5.8 (0.04) & 13 (41.08) & 8 (0.04) & 7.2 (8.29) & 7.4 (4.05) \\
    \midrule
    \multirow{4}{*}{$\left(
                      \begin{array}{cc}
                        0.05 & 0.02 \\
                        0.02 & 0.05 \\
                      \end{array}
                    \right)$
                    }   & 10  & 3   & 25.5 $(<0.01)$ & 25.4 (0.04) & 34.7 (38.66) & 47.9 (0.03) & 23.7 (7.94) & 23.8 (3.86) \\
       & 10  & 5   & 81.7 $(<0.01)$ & 74.5 (0.05) & 86.3 (39.07) & 90.7 (0.03) & 79.9 (8.02) & 80.2 (3.93) \\
       & 15  & 3   & 35.8 $(<0.01)$ & 38.3 (0.05) & 52.2 (39.86) & 60.9 (0.03) & 30.3 (8.15) & 30.8 (3.99) \\
       & 15  & 5   & 89.7 $(<0.01)$ & 87.2 (0.05) & 93.6 (41.04) & 95.2 (0.04) & 88.1 (8.16) & 88.4 (4.07) \\
    \midrule
    \multirow{4}{*}{$\left(
                      \begin{array}{cc}
                        0.08 & 0.02 \\
                        0.02 & 0.08 \\
                      \end{array}
                    \right)$
                    }   & 10  & 3   & 34.5 $(<0.01)$ & 35.6 (0.04) & 48.9 (38.66) & 60.9 (0.03) & 32.1 (7.78) & 33.1 (3.87) \\
       & 10  & 5   & 90.3 $(<0.01)$ & 86.7 (0.05) & 94.1 (39.08) & 95.7 (0.03) & 89.2 (7.91) & 89.8 (3.94) \\
       & 15  & 3   & 46.9 $(<0.01)$ & 53.5 (0.05) & 66.5 (39.85) & 73.9 (0.03) & 42.9 (8.02) & 43 (4) \\
       & 15  & 5   & 95.5 $(<0.01)$ & 94.9 (0.05) & 97.9 (41.03) & 98.5 (0.04) & 94.6 (8.18) & 94.9 (4.16) \\
    \midrule
    \multirow{4}{*}{$\left(
                      \begin{array}{cc}
                        0.1 & 0.05 \\
                        0.05 & 0.1 \\
                      \end{array}
                    \right)$
                    }   & 10  & 3   & 44.9 $(<0.01)$ & 50.1 (0.04) & 63.1 (38.68) & 75.1 (0.03) & 42 (7.79) & 43.8 (4) \\
       & 10  & 5   & 95.1 $(<0.01)$ & 93.5 (0.05) & 97 (39.09) & 98.2 (0.03) & 94.1 (8.03) & 94.3 (3.94) \\
       & 15  & 3   & 61.3 $(<0.01)$ & 69.2 (0.05) & 78.5 (39.84) & 86.1 (0.03) & 57.8 (8.17) & 58.2 (3.99) \\
       & 15  & 5   & 98.2 $(<0.01)$ & 97.9 (0.05) & 99 (41.04) & 99.4 (0.04) & 97.5 (8.17) & 97.7 (4.06) \\
    \bottomrule
    \end{tabular}}%
  \label{tab:s1-student}%
\end{table}%

\begin{table}[H]
  \centering
  \caption{Simulation results for Setting $1$ (Table 2 in the original manuscript) for cases with non-normal error with chi-squared distribution with 3 degrees of freedom. Methods compared included the FLC test, the standard likelihood ratio test (LRT) compared to a Chi-squared test, its parametric bootstrap (PB) counterpart, the linear score test (LinScore), bootstrap FLC test using fast double bootstrap (FDB) and residual bootstrap FLC test (BT). $D$ is the variance-covariance matrix for variance components. $m$ and $n$ are the number of clusters and cluster size, respectively. Performance was assessed in terms of percentage of datasets where the method rejected the null hypothesis, and the mean computation time in seconds(in parentheses).}
  \resizebox{\textwidth}{!}{%
        \begin{tabular}{lllllllll}
    \toprule
    \textbf{D}   & n   & m   & FLC & LRT & PB  & LinScore & FDB & BT \\
    \midrule
       \multirow{4}{*}{$\left(
                      \begin{array}{cc}
                        0 & 0 \\
                        0 & 0 \\
                      \end{array}
                    \right)$}   & 10  & 3   & 6.5 $(<0.01)$ & 5.8 (0.04) & 10.4 (38.69) & 5.4 (0.03) & 5.5 (7.94) & 5.4 (3.88) \\
       & 10  & 5   & 4.9 $(<0.01)$ & 2.6 (0.04) & 7.9 (39.07) & 4 (0.03) & 4.9 (7.91) & 4.4 (3.94) \\
     & 15  & 3   & 7.3 $(<0.01)$ & 5.8 (0.04) & 11.3 (39.83) & 5.5 (0.03) & 4.8 (8.15) & 5.6 (3.98) \\
       & 15  & 5   & 5.9 $(<0.01)$ & 2.1 (0.04) & 7.8 (41.05) & 5.8 (0.04) & 4.6 (8.16) & 4.8 (4.06) \\
    \midrule
        \multirow{4}{*}{$\left(
                      \begin{array}{cc}
                        0.05 & 0.02 \\
                        0.02 & 0.05 \\
                      \end{array}
                    \right)$}   & 10  & 3   & 21.8 $(<0.01)$ & 18.7 (0.04) & 28.8 (38.69) & 40.3 (0.03) & 21.5 (7.79) & 21 (3.88) \\
       & 10  & 5   & 75.7 $(<0.01)$ & 66.2 (0.05) & 81.6 (39.09) & 88 (0.03) & 74.4 (7.93) & 74.6 (4.04) \\
       & 15  & 3   & 27.2 $(<0.01)$ & 27.1 (0.04) & 39.9 (39.82) & 52.6 (0.03) & 25 (8.03) & 24.8 (4.01) \\
       & 15  & 5   & 87.6 $(<0.01)$ & 84.5 (0.05) & 92.5 (41.02) & 96.4 (0.04) & 85.4 (8.28) & 86.8 (4.05) \\
    \midrule
        \multirow{4}{*}{$\left(
                      \begin{array}{cc}
                        0.08 & 0.02 \\
                        0.02 & 0.08 \\
                      \end{array}
                    \right)$}   & 10  & 3   & 29.6 $(<0.01)$ & 28.3 (0.04) & 41.9 (38.69) & 54.6 (0.03) & 27.4 (7.93) & 28 (3.86) \\
       & 10  & 5   & 88.2 $(<0.01)$ & 83.5 (0.05) & 92 (39.09) & 94.5 (0.03) & 88.3 (8.03) & 87.7 (3.92) \\
       & 15  & 3   & 38.7 $(<0.01)$ & 42.4 (0.05) & 55.1 (39.81) & 66.8 (0.03) & 36 (8.15) & 35.5 (3.99) \\
       & 15  & 5   & 96.5 $(<0.01)$ & 95.2 (0.05) & 98.3 (41.01) & 99 (0.04) & 95.8 (8.16) & 96 (4.06) \\
    \midrule
        \multirow{4}{*}{$\left(
                      \begin{array}{cc}
                        0.1 & 0.05 \\
                        0.05 & 0.1 \\
                      \end{array}
                    \right)$}   & 10  & 3   & 40.6 $(<0.01)$ & 43.5 (0.04) & 55.1 (38.69) & 68.4 (0.03) & 38.8 (7.78) & 39 (3.88) \\
       & 10  & 5   & 94 $(<0.01)$ & 91.7 (0.05) & 96.4 (39.09) & 97.9 (0.03) & 93.5 (7.91) & 94 (3.94) \\
       & 15  & 3   & 53 $(<0.01)$ & 59.1 (0.05) & 70.9 (39.8) & 81.6 (0.03) & 49.5 (8.03) & 51.3 (4) \\
       & 15  & 5   & 98.8 $(<0.01)$ & 98.6 (0.05) & 99.5 (41.03) & 99.8 (0.04) & 98.6 (8.29) & 98.5 (4.05) \\
    \bottomrule
    \end{tabular}}%
  \label{tab:s1-chi}
\end{table}%

\begin{table}[H]
  \centering
  \caption{Simulation results for Setting $1$ (Table 2 in the original manuscript) for cases with contamination. Methods compared included the FLC test, the standard likelihood ratio test (LRT) compared to a Chi-squared test, its parametric bootstrap (PB) counterpart, the linear score test (LinScore), bootstrap FLC test using fast double bootstrap (FDB) and residual bootstrap FLC test (BT). $D$ is the variance-covariance matrix for variance components. $m$ and $n$ are the number of clusters and cluster size, respectively. Performance was assessed in terms of percentage of datasets where the method rejected the null hypothesis, and the mean computation time in seconds(in parentheses).}
    \resizebox{\textwidth}{!}{%
    \begin{tabular}{lllllllll}
    \toprule
    \textbf{D}   & n   & m   & FLC & LRT & PB  & LinScore & FDB & BT \\
    \midrule
        \multirow{4}{*}{$\left(
                      \begin{array}{cc}
                        0 & 0 \\
                        0 & 0 \\
                      \end{array}
                    \right)$}   & 10  & 3   & 5.4 $(<0.01)$ & 5.9 (0.04) & 12 (38.64) & 5.8 (0.03) & 3.8 (7.8) & 4.2 (4.02) \\
       & 10  & 5   & 4.8 $(<0.01)$ & 2.2 (0.04) & 7.1 (39.08) & 3.7 (0.03) & 4.7 (8.04) & 4.5 (3.93) \\
       & 15  & 3   & 9.1 $(<0.01)$ & 7.3 (0.04) & 13.3 (39.83) & 5.4 (0.03) & 7 (8.04) & 6.7 (4.11) \\
       & 15  & 5   & 6.3 $(<0.01)$ & 2.3 (0.04) & 8.6 (41.06) & 5.4 (0.04) & 4.5 (8.27) & 4.8 (4.05) \\
    \midrule
    \multirow{4}{*}{$\left(
                      \begin{array}{cc}
                        0.05 & 0.02 \\
                        0.02 & 0.05 \\
                      \end{array}
                    \right)$}     & 10  & 3   & 23.1 $(<0.01)$ & 18.9 (0.04) & 29.9 (38.64) & 40.6 (0.03) & 19.1 (7.91) & 20.1 (3.87) \\
       & 10  & 5   & 76.1 $(<0.01)$ & 66.4 (0.04) & 81.9 (39.09) & 88.6 (0.03) & 74.6 (7.91) & 74.2 (3.94) \\
       & 15  & 3   & 29.9 $(<0.01)$ & 27.5 (0.04) & 40.4 (39.83) & 51 (0.03) & 27.1 (8.14) & 27.6 (3.99) \\
       & 15  & 5   & 87.8 $(<0.01)$ & 84.9 (0.05) & 92.3 (41.02) & 94.6 (0.04) & 86.1 (8.17) & 86.2 (4.07) \\
    \midrule
    \multirow{4}{*}{$\left(
                      \begin{array}{cc}
                        0.08 & 0.02 \\
                        0.02 & 0.08 \\
                      \end{array}
                    \right)$}     & 10  & 3   & 30.6 $(<0.01)$ & 28.5 (0.04) & 41.1 (38.66) & 54.4 (0.03) & 28.2 (7.8) & 29.6 (3.87) \\
       & 10  & 5   & 89.5 $(<0.01)$ & 85.1 (0.05) & 92.4 (39.09) & 96.6 (0.03) & 89.1 (7.93) & 88.3 (4.04) \\
       & 15  & 3   & 39.6 $(<0.01)$ & 42.6 (0.04) & 55.8 (39.81) & 68.2 (0.03) & 36.1 (8.03) & 36.9 (4.12) \\
       & 15  & 5   & 95.7 $(<0.01)$ & 94.4 (0.06) & 97.5 (41.05) & 98.6 (0.04) & 95.5 (8.28) & 95.5 (4.06) \\
    \midrule
    \multirow{4}{*}{$\left(
                      \begin{array}{cc}
                        0.1 & 0.05 \\
                        0.05 & 0.1 \\
                      \end{array}
                    \right)$}     & 10  & 3   & 41.3 $(<0.01)$ & 41.5 (0.04) & 55.8 (38.66) & 69.5 (0.03) & 39.2 (7.92) & 39.2 (3.86) \\
       & 10  & 5   & 95 $(<0.01)$ & 92.9 (0.05) & 97.3 (39.09) & 98.5 (0.03) & 94.9 (8.02) & 94.8 (3.93) \\
       & 15  & 3   & 51.9 $(<0.01)$ & 60.6 (0.05) & 72.3 (39.81) & 82.9 (0.03) & 48.9 (8.15) & 49.4 (3.99) \\
       & 15  & 5   & 98.2 $(<0.01)$ & 98 (0.05) & 99.1 (41.06) & 99.6 (0.04) & 98 (8.16) & 98.2 (4.06) \\
    \bottomrule
    \end{tabular}}%
  \label{tab:s1-contam}%
\end{table}%

\begin{table}[H]
  \centering
  \caption{Simulation results for Setting $2$ (Table 3 in the original manuscript) for cases with Student's t distribution with 3 degrees of freedom. Methods compared included the FLC test, the parametric bootstrap likelihood ratio test (PB), the linear score test (LinScore), bootstrap FLC test using fast double bootstrap (FDB) and residual bootstrap FLC test (BT). $D$ is the variance-covariance matrix for variance components, and $n$ is the cluster size. Performance was assessed in terms of percentage of datasets where the method rejected the null hypothesis, and the mean computation time in seconds(in parentheses).}
    \begin{tabular}{rrlllll}
    \toprule
   \textbf{D} & n & FLC & PB  & LinScore & FDB & BT \\
    \midrule
      \multirow{4}{*}{$\left(
                      \begin{array}{cc}
                        0 & 0 \\
                        0 & 0 \\
                      \end{array}
                    \right)$}     & 7   & 6.3 (0.02) & 6.8 (37.44) & 5 (0.09) & 4.6 (8.66) & 4.4 (4.41) \\
       & 15  & 9 $(<0.01)$ & 10.1 (41.81) & 7.4 (0.29) & 6.2 (10.35) & 7 (5.09) \\
       & 25  & 11.5 $(<0.01)$ & 15.1 (46.43) & 8.7 (0.71) & 7 (15.9) & 7.8 (7.47) \\
       & 50  & 9.8 (0.02) & 16.2 (58) & 9.8 (4.19) & 5.2 (47.94) & 6.2 (18.04) \\
    \midrule
      \multirow{4}{*}{$\left(
                      \begin{array}{cc}
                        0.2 & 0.1 \\
                        0.1 & 0.2 \\
                      \end{array}
                    \right)$}     & 7   & 11.9 $(<0.01)$ & 10.4 (39.27) & 10.8 (0.08) & 10.1 (8.61) & 10.7 (4.29) \\
       & 15  & 18.2 $(<0.01)$ & 18.6 (43.45) & 18.1 (0.21) & 14.5 (10.35) & 14.5 (5.07) \\
       & 25  & 25.2 $(<0.01)$ & 25.2 (48.16) & 25.6 (0.56) & 19 (15.83) & 19.1 (7.48) \\
       & 50  & 27.5 (0.02) & 30.7 (60.73) & 31 (3.77) & 17.9 (47.59) & 19.9 (18.16) \\
    \midrule
      \multirow{4}{*}{$\left(
                      \begin{array}{cc}
                        0.5 & 0.1 \\
                        0.1 & 0.5 \\
                      \end{array}
                    \right)$}     & 7   & 24.1 $(<0.01)$ & 21.5 (39.4) & 25.5 (0.07) & 21.6 (8.73) & 21.8 (4.28) \\
       & 15  & 37.6 $(<0.01)$ & 36.6 (43.69) & 40.4 (0.2) & 32.1 (10.36) & 33.4 (5.08) \\
       & 25  & 52.9 $(<0.01)$ & 49.1 (48.62) & 54.9 (0.53) & 44.5 (15.97) & 47 (7.49) \\
       & 50  & 65.4 (0.02) & 63.6 (61.94) & 69.5 (3.51) & 53.2 (47.53) & 56 (18.2) \\
    \midrule
      \multirow{4}{*}{$\left(
                      \begin{array}{cc}
                        1 & 0.2 \\
                        0.2 & 0.1 \\
                      \end{array}
                    \right)$}     & 7   & 46.5 $(<0.01)$ & 42.2 (39.53) & 47.4 (0.07) & 42.6 (8.6) & 43.9 (4.28) \\
       & 15  & 66.2 $(<0.01)$ & 65.8 (44.06) & 70.7 (0.19) & 60.2 (10.35) & 62.4 (5.08) \\
       & 25  & 80 $(<0.01)$ & 79.3 (49.01) & 83.5 (0.53) & 74.7 (16.04) & 75.7 (7.5) \\
       & 50  & 91.9 (0.02) & 91.7 (62.8) & 93 (3.44) & 84.8 (47.61) & 86.1 (18.22) \\
    \bottomrule
    \end{tabular}
  \label{tab:s2-student}%
\end{table}%

\begin{table}[H]
  \centering
  \caption{Simulation results for Setting $2$ (Table 3 in the original manuscript) for cases with chi-squared distribution with 3 degrees of freedom. Methods compared included the FLC test, the parametric bootstrap likelihood ratio test (PB), the linear score test (LinScore), bootstrap FLC test using fast double bootstrap (FDB) and residual bootstrap FLC test (BT). $D$ is the variance-covariance matrix for variance components, and $n$ is the cluster size. Performance was assessed in terms of percentage of datasets where the method rejected the null hypothesis, and the mean computation time in seconds(in parentheses).}
   \resizebox{\textwidth}{!}{%
        \begin{tabular}{rrlllll}
    \toprule
    \textbf{D} & n & FLC & PB  & LinScore & FDB & BT \\
    \midrule
          \multirow{4}{*}{$\left(
                      \begin{array}{cc}
                        0 & 0 \\
                        0 & 0 \\
                      \end{array}
                    \right)$}      & 7   & 5.9 $(<0.01)$ & 5.8 (37.63) & 6.1 (0.07) & 4.7 (8.63) & 5.4 (4.3) \\
       & 15  & 4.8 $(<0.01)$ & 5.6 (42.03) & 4.5 (0.23) & 3.5 (10.35) & 3.8 (5.09) \\
       & 25  & 7 $(<0.01)$ & 8.1 (46.52) & 6.4 (0.56) & 5.8 (15.8) & 5.7 (7.51) \\
       & 50  & 6.5 (0.02) & 8.3 (58.67) & 5.8 (3.72) & 4.8 (47.62) & 4.9 (18.14) \\
    \midrule
              \multirow{4}{*}{$\left(
                      \begin{array}{cc}
                        0.2 & 0.1 \\
                        0.1 & 0.2 \\
                      \end{array}
                    \right)$}     & 7   & 11.9 $(<0.01)$ & 11.9 (39.15) & 10.9 (0.08) & 10 (8.73) & 10.4 (4.29) \\
       & 15  & 13.5 $(<0.01)$ & 12.7 (43.37) & 14.7 (0.21) & 11.8 (10.36) & 11.6 (5.08) \\
       & 25  & 19.5 $(<0.01)$ & 18.8 (47.99) & 20 (0.55) & 16.9 (15.83) & 17.7 (7.47) \\
       & 50  & 23.9 (0.02) & 23.7 (60.91) & 26.8 (3.62) & 20.4 (47.67) & 21.2 (18.08) \\
    \midrule
              \multirow{4}{*}{$\left(
                      \begin{array}{cc}
                        0.5 & 0.1 \\
                        0.1 & 0.5 \\
                      \end{array}
                    \right)$}     & 7   & 21.3 $(<0.01)$ & 20.3 (39.25) & 23.4 (0.07) & 20.4 (8.61) & 19.9 (4.29) \\
       & 15  & 29.7 $(<0.01)$ & 27.5 (43.6) & 34.1 (0.21) & 27.3 (10.35) & 27.9 (5.1) \\
       & 25  & 44.7 $(<0.01)$ & 42.2 (48.43) & 47.7 (0.52) & 41.3 (15.96) & 41.9 (7.58) \\
       & 50  & 63.7 (0.02) & 59.9 (61.92) & 68.4 (3.48) & 57.1 (47.4) & 57.9 (18.11) \\
    \midrule
              \multirow{4}{*}{$\left(
                      \begin{array}{cc}
                        1 & 0.2 \\
                        0.2 & 0.1 \\
                      \end{array}
                    \right)$}     & 7   & 39.3 $(<0.01)$ & 37.2 (39.38) & 40.2 (0.07) & 37.5 (8.73) & 37.8 (4.28) \\
       & 15  & 59.5 $(<0.01)$ & 58.2 (43.87) & 64.1 (0.19) & 56.2 (10.35) & 55.9 (5.09) \\
       & 25  & 77.7 $(<0.01)$ & 73.8 (48.95) & 80.8 (0.52) & 74.5 (15.96) & 75.9 (7.57) \\
       & 50  & 95 (0.02) & 94 (62.88) & 97.5 (3.41) & 93.8 (47.72) & 94.4 (18.24) \\
    \bottomrule
    \end{tabular}}%
  \label{tab:s2-chi}%
\end{table}%

\begin{table}[H]
  \centering
  \caption{Simulation results for Setting $2$ (Table 3 in the original manuscript) for cases with contamination. Methods compared included the FLC test, the parametric bootstrap likelihood ratio test (PB), the linear score test (LinScore), bootstrap FLC test using fast double bootstrap (FDB) and residual bootstrap FLC test (BT). $D$ is the variance-covariance matrix for variance components, and $n$ is the cluster size. Performance was assessed in terms of percentage of datasets where the method rejected the null hypothesis, and the mean computation time in seconds(in parentheses).}
       \resizebox{\textwidth}{!}{%
    \begin{tabular}{rrlllll}
    \toprule
   \textbf{D} & n & FLC & PB  & LinScore & FDB & BT \\
    \midrule
              \multirow{4}{*}{$\left(
                      \begin{array}{cc}
                        0 & 0 \\
                        0 & 0 \\
                      \end{array}
                    \right)$}     & 7   & 5 $(<0.01)$ & 6.1 (37.58) & 6.2 (0.07) & 3.9 (8.74) & 4.6 (4.28) \\
       & 15  & 6.6 $(<0.01)$ & 7.1 (41.87) & 6.1 (0.2) & 5 (10.36) & 5.4 (5.09) \\
       & 25  & 5.8 $(<0.01)$ & 6.7 (46.38) & 5.6 (0.56) & 4.9 (15.82) & 4.6 (7.48) \\
       & 50  & 7 (0.02) & 7.3 (58.41) & 5.2 (3.78) & 5.4 (47.71) & 5.4 (18.06) \\
    \midrule
              \multirow{4}{*}{$\left(
                      \begin{array}{cc}
                        0.2 & 0.1 \\
                        0.1 & 0.2 \\
                      \end{array}
                    \right)$}     & 7   & 9.8 $(<0.01)$ & 10.4 (39.15) & 10.7 (0.08) & 8.1 (8.62) & 8.9 (4.29) \\
       & 15  & 14.5 $(<0.01)$ & 15.2 (43.35) & 15.3 (0.21) & 11.5 (10.36) & 12.1 (5.09) \\
       & 25  & 17.1 $(<0.01)$ & 18.9 (48.05) & 19 (0.55) & 13.7 (15.96) & 14.5 (7.48) \\
       & 50  & 25.3 (0.02) & 24.2 (60.95) & 29.2 (3.59) & 20.4 (47.86) & 21.3 (18.09) \\
    \midrule
              \multirow{4}{*}{$\left(
                      \begin{array}{cc}
                        0.5 & 0.1 \\
                        0.1 & 0.5 \\
                      \end{array}
                    \right)$}     & 7   & 19.5 $(<0.01)$ & 19.8 (39.26) & 21.9 (0.07) & 17.2 (8.72) & 18.9 (4.27) \\
       & 15  & 31.1 $(<0.01)$ & 30.9 (43.64) & 34.5 (0.19) & 28.2 (10.36) & 27.9 (5.18) \\
       & 25  & 40.8 $(<0.01)$ & 38.7 (48.53) & 45.4 (0.52) & 35.8 (15.96) & 36.1 (7.51) \\
       & 50  & 61.9 (0.02) & 59.8 (62.09) & 67.5 (3.46) & 57.1 (47.43) & 57.7 (18.22) \\
    \midrule
              \multirow{4}{*}{$\left(
                      \begin{array}{cc}
                        1 & 0.2 \\
                        0.2 & 0.1 \\
                      \end{array}
                    \right)$}     & 7   & 38.6 $(<0.01)$ & 36.7 (39.41) & 39.7 (0.07) & 36.1 (8.62) & 37.6 (4.39) \\
       & 15  & 60.9 $(<0.01)$ & 56.2 (43.91) & 62.1 (0.18) & 56.2 (10.35) & 58.1 (5.1) \\
       & 25  & 77.3 $(<0.01)$ & 74.1 (49.08) & 80.3 (0.51) & 72.6 (15.96) & 73.9 (7.51) \\
       & 50  & 95 (0.02) & 92.8 (62.96) & 97.3 (3.44) & 93.5 (47.69) & 94.2 (18.24) \\
    \bottomrule
    \end{tabular}}%
  \label{tab:s2-contam}%
\end{table}%

\begin{table}[H]
  \centering
  \caption{Simulation results for Setting $3$ (Table 4 in the original manuscript) for cases with Student's t distribution with 3 degrees of freedom. Methods compared included the FLC test, the parametric bootstrap likelihood ratio test (PB), the linear score test (LinScore), bootstrap FLC test using fast double bootstrap (FDB) and residual bootstrap FLC test (BT). $\tau$ is diagonal value for the variance-covariance matrix for variance components, and the off diagonal values are $\tau/2$. $m$ and $n$ are the number of clusters and cluster size, respectively. Performance was assessed in terms of percentage of datasets where the method rejected the null hypothesis, and the mean computation time in seconds(in parentheses).}
  \begin{tabular}{rrrlllll}
    \toprule
    \multicolumn{1}{l}{$\tau$} & n & m & FLC & PB  & LinScore & FDB & BT \\
    \midrule
     \multirow{6}{*}{0}  & 10  & 10  & 7.4 (0.02) & 7.3 (136.5) & 5 (0.33) & 5.8 (9.97) & 6.3 (4.94) \\
       & 10  & 20  & 7.2 $(<0.01)$ & 8.2 (186.9) & 5.4 (0.86) & 4.8 (10.83) & 5.3 (5.37) \\
       & 20  & 10  & 11.7 $(<0.01)$ & 11.8 (178.28) & 7 (1.1) & 7.4 (16.13) & 7.7 (7.5) \\
     & 20  & 20  & 7.3 $(<0.01)$ & 9.3 (276.17) & 5.4 (5.43) & 5 (19.36) & 5.3 (9.14) \\
       & 40  & 10  & 11.4 (0.01) & 13.1 (264.99) & 6.3 (5.52) & 6.3 (44.86) & 7.2 (17.68) \\
       & 40  & 20  & 10.8 (0.02) & 13.2 (459) & 6.5 (38.53) & 5.5 (99.81) & 6 (42.31) \\
    \midrule
    \multirow{6}{*}{0.1}   & 10  & 10  & 48.6 $(<0.01)$ & 49 (134.69) & 53.3 (0.2) & 44.3 (9.95) & 46.8 (4.83) \\
       & 10  & 20  & 84.1 $(<0.01)$ & 79.6 (183.98) & 84.9 (0.76) & 80.6 (10.82) & 81.3 (5.28) \\
       & 20  & 10  & 71.1 $(<0.01)$ & 73.4 (176.35) & 77 (0.77) & 64.5 (16.1) & 66.2 (7.59) \\
       & 20  & 20  & 97.2 (0.01) & 96.9 (276.79) & 97.4 (4.51) & 93.9 (19.39) & 94 (9.21) \\
       & 40  & 10  & 89.1 (0.01) & 92.3 (268.72) & 93 (4.29) & 80.2 (44.85) & 83.3 (17.72) \\
       & 40  & 20  & 99 (0.02) & 99.6 (475.62) & 99.3 (31.19) & 97.3 (99.74) & 97.8 (42.32) \\
    \midrule
    \multirow{6}{*}{0.2}   & 10  & 10  & 77 $(<0.01)$ & 75.9 (133.29) & 79 (0.2) & 73.2 (9.94) & 74.8 (4.85) \\
       & 10  & 20  & 96.5 $(<0.01)$ & 95.4 (181.23) & 96.7 (0.75) & 95.3 (10.82) & 95.4 (5.35) \\
       & 20  & 10  & 92.8 $(<0.01)$ & 94.3 (174.18) & 95.4 (0.74) & 88.1 (16.09) & 89.7 (7.58) \\
       & 20  & 20  & 99.7 (0.01) & 99.8 (272.91) & 99.8 (4.43) & 99 (19.38) & 99.2 (9.13) \\
       & 40  & 10  & 98.3 (0.02) & 99.4 (265.83) & 99 (4.2) & 94.8 (44.8) & 96.1 (17.75) \\
       & 40  & 20  & 99.6 (0.02) & 99.8 (464.01) & 99.7 (30.53) & 98.8 (99.72) & 99 (42.34) \\
    \bottomrule
      \end{tabular}
  \label{tab:s3-student}%
\end{table}%

\begin{table}[H]
  \centering
  \caption{Simulation results for Setting $3$ (Table 4 in the original manuscript) for cases with chi-squared distribution with 3 degrees of freedom. Methods compared included the FLC test, the parametric bootstrap likelihood ratio test (PB), the linear score test (LinScore), bootstrap FLC test using fast double bootstrap (FDB) and residual bootstrap FLC test (BT). $\tau$ is diagonal value for the variance-covariance matrix for variance components, and the off diagonal values are $\tau/2$. $m$ and $n$ are the number of clusters and cluster size, respectively. Performance was assessed in terms of percentage of datasets where the method rejected the null hypothesis, and the mean computation time in seconds(in parentheses).}
      \resizebox{\textwidth}{!}{%
        \begin{tabular}{rrrlllll}
    \toprule
    \multicolumn{1}{l}{$\tau$} & n & m & FLC & PB  & LinScore & FDB & BT \\
    \midrule
    \multirow{6}{*}{0}   & 10  & 10  & 6.2 $(<0.01)$ & 7.5 (134.74) & 3.9 (0.22) & 5.3 (9.96) & 5.5 (4.96) \\
       & 10  & 20  & 7.6 $(<0.01)$ & 7.9 (183.72) & 4 (0.85) & 6.5 (10.83) & 6.9 (5.27) \\
       & 20  & 10  & 9.5 $(<0.01)$ & 8.7 (175.8) & 4.7 (0.89) & 7.4 (16.22) & 7.1 (7.49) \\
       & 20  & 20  & 6.6 $(<0.01)$ & 6.8 (273.19) & 4.2 (5.27) & 5.4 (19.38) & 6.3 (9.13) \\
       & 40  & 10  & 7.6 (0.02) & 8.2 (262.3) & 5.1 (4.96) & 6.1 (44.89) & 5.4 (17.69) \\
       & 40  & 20  & 6.3 (0.02) & 6.3 (454.92) & 3.7 (37.17) & 5.4 (100.03) & 5.3 (42.43) \\
    \midrule
    \multirow{6}{*}{0.1}   & 10  & 10  & 41.4 $(<0.01)$ & 41.7 (132.88) & 46.3 (0.2) & 38.7 (9.96) & 39.7 (4.92) \\
       & 10  & 20  & 85.7 $(<0.01)$ & 80.1 (181.23) & 84.7 (0.75) & 84 (10.81) & 84.6 (5.35) \\
       & 20  & 10  & 71.3 $(<0.01)$ & 69.8 (174.26) & 78.6 (0.73) & 66.7 (16.11) & 68.4 (7.5) \\
       & 20  & 20  & 98.6 $(<0.01)$ & 97.2 (274.05) & 98.1 (4.51) & 97.5 (19.32) & 97.9 (9.22) \\
       & 40  & 10  & 90 (0.01) & 92 (266.01) & 94.5 (4.22) & 87 (44.81) & 87.6 (17.73) \\
       & 40  & 20  & 99.9 (0.02) & 99.8 (469.88) & 99.9 (30.69) & 99.9 (99.65) & 99.9 (42.33) \\
    \midrule
    \multirow{6}{*}{0.2}   & 10  & 10  & 71.4 $(<0.01)$ & 71.5 (131.78) & 75.8 (0.2) & 70 (10.03) & 70.5 (4.83) \\
       & 10  & 20  & 97.5 $(<0.01)$ & 95.9 (178.89) & 97 (0.74) & 97.3 (10.83) & 97.1 (5.29) \\
       & 20  & 10  & 94.8 $(<0.01)$ & 93.8 (172.47) & 97.1 (0.73) & 94.1 (16.08) & 94.1 (7.51) \\
       & 20  & 20  & 100 $(<0.01)$ & 100 (270.5) & 100 (4.42) & 100 (19.37) & 100 (9.12) \\
      & 40  & 10  & 99.7 (0.01) & 99.6 (264.05) & 99.7 (4.12) & 99.2 (44.8) & 99.3 (17.77) \\
       & 40  & 20  & 100 (0.02) & 100 (457.7) & 100 (29.87) & 100 (99.8) & 100 (42.45) \\
    \bottomrule
    \end{tabular}}%
  \label{tab:s3-chi}%
\end{table}%

\begin{table}[H]
  \centering
  \caption{Simulation results for Setting $3$ (Table 4 in the original manuscript) for cases with contamination. Methods compared included the FLC test, the parametric bootstrap likelihood ratio test (PB), the linear score test (LinScore), bootstrap FLC test using fast double bootstrap (FDB) and residual bootstrap FLC test (BT). $\tau$ is diagonal value for the variance-covariance matrix for variance components, and the off diagonal values are $\tau/2$. $m$ and $n$ are the number of clusters and cluster size, respectively. Performance was assessed in terms of percentage of datasets where the method rejected the null hypothesis, and the mean computation time in seconds(in parentheses).}
      \resizebox{\textwidth}{!}{%
        \begin{tabular}{rrrlllll}
    \toprule
    \multicolumn{1}{l}{$\tau$} & n & m & FLC & PB  & LinScore & FDB & BT \\
    \midrule
     \multirow{6}{*}{0}   & 10  & 10  & 7.1 $(<0.01)$ & 7.9 (134.54) & 3.9 (0.22) & 5.9 (9.96) & 5.8 (4.97) \\
       & 10  & 20  & 6.4 $(<0.01)$ & 6.6 (183.8) & 3.9 (0.85) & 5.4 (10.83) & 5.5 (5.35) \\
       & 20  & 10  & 7 $(<0.01)$ & 8.2 (175.76) & 5.7 (0.87) & 5.3 (16.1) & 5.4 (7.5) \\
       & 20  & 20  & 6.3 $(<0.01)$ & 6.3 (273.61) & 4.1 (5.25) & 5.4 (19.37) & 5 (9.21) \\
       & 40  & 10  & 7.6 (0.02) & 7.3 (263.43) & 5.3 (4.92) & 6.1 (44.87) & 6.5 (17.69) \\
       & 40  & 20  & 6.3 (0.02) & 5.4 (453.82) & 4.1 (36.86) & 4.6 (99.97) & 4.7 (42.33) \\
    \midrule
    \multirow{6}{*}{0.1}    & 10  & 10  & 46.9 $(<0.01)$ & 45.2 (132.67) & 50.6 (0.2) & 42.6 (9.94) & 44.1 (4.93) \\
       & 10  & 20  & 85.7 $(<0.01)$ & 79.9 (181.17) & 85.7 (0.75) & 83.3 (10.82) & 84 (5.36) \\
       & 20  & 10  & 71 $(<0.01)$ & 69 (174.17) & 77.9 (0.74) & 65 (16.1) & 67.8 (7.59) \\
       & 20  & 20  & 97.9 $(<0.01)$ & 96.4 (274.17) & 98.1 (4.51) & 97.3 (19.37) & 97.7 (9.13) \\
       & 40  & 10  & 91.4 (0.01) & 92.3 (266.36) & 95.2 (4.21) & 87.7 (44.81) & 89 (17.74) \\
       & 40  & 20  & 100 (0.02) & 100 (468.8) & 100 (30.56) & 100 (99.82) & 100 (42.25) \\
    \midrule
    \multirow{6}{*}{0.2}    & 10  & 10  & 74.7 $(<0.01)$ & 73.5 (131.54) & 76.3 (0.19) & 72.3 (9.95) & 72.7 (4.91) \\
       & 10  & 20  & 97.3 $(<0.01)$ & 95.6 (178.89) & 97.2 (0.75) & 97 (10.83) & 97.2 (5.35) \\
       & 20  & 10  & 94.7 $(<0.01)$ & 95.2 (172.21) & 96.2 (0.72) & 92.8 (16.09) & 93.9 (7.58) \\
       & 20  & 20  & 100 $(<0.01)$ & 99.8 (270.95) & 99.8 (4.41) & 99.9 (19.25) & 99.9 (9.2) \\
       & 40  & 10  & 99.8 (0.01) & 99.6 (264) & 99.9 (4.14) & 99.8 (44.8) & 99.7 (17.77) \\
       & 40  & 20  & 100 (0.02) & 100 (454.75) & 100 (29.74) & 100 (99.81) & 100 (42.36) \\
    \bottomrule
    \end{tabular}}%
  \label{tab:s3-contam}%
\end{table}%

\newpage

\section{Notes on Comparing Double Bootstrap and Fast Double bootstrap}
\label{appdx:doubleBT}

The algorithm for double bootstrapping the FLC test is as follows:
\begin{enumerate}
  \item Obtain the test statistic $F_{obs}$.
  \item Generate $B_1$ bootstrap samples which are constructed under the null hypothesis, and compute a bootstrap statistic $F^*_k$ for each $k=1,\ldots,B_1$ bootstrap sample.
  \item Calculate the first-level bootstrap P value as $\hat{p}^*=\frac{1}{B_1} \sum_{k=1}^{B_1} I(F^*_k > F_{obs})$.
  \item For each of the $B_1$ first-level bootstrap samples, generate $B_2$ second-level bootstrap samples. Use the second-level bootstrap samples to compute a second-level bootstrap test statistic $F^{**}_{kl}$ for $l=1,\ldots,B_2$.
  \item For each of the $B_1$ first-level bootstrap samples, compute the second-level bootstrap P value $\hat{p}^{**}_k=\frac{1}{B_2} \sum_{l=1}^{B_2} I(F^{**}_{kl} > F^*_{k})$.
  \item Calculate the \emph{double bootstrap} P value as
  \begin{equation*}
  \hat{p}^{**}= \frac{1}{B_1} \sum^{B_1}_{k=1} I ( \hat{p}^{**}_k < \hat{p}^* )\,.
    \end{equation*}
\end{enumerate}

\begin{figure}[H]
 \begin{center}	
\includegraphics[width=17cm]{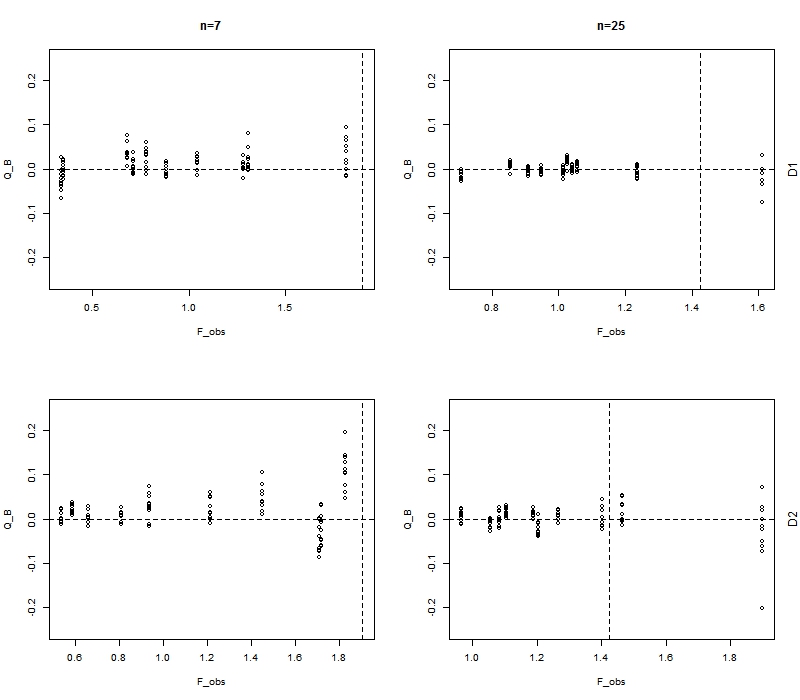}
 \end{center}
\caption{Plots of fast double bootstrap $1-\hat{p}^*$ quantile ($\hat{Q}_{B}^{**}$) aginst statistics for FLC test ($F_{obs}$) for $10$ datasets generated using the Setting $2$ with Student's t distribution with 3 degrees of freedom, when the size of the cluster $m=10$. Each row represents one of two covariance matrices $\mathbf{D}_{1}=$~\usebox{\smlmatit} and $\mathbf{D}_{2}=$~\usebox{\smlmatiit}. The values on the y axis show the differences between the $\hat{Q}_{B}^{**}$ and their corresponding FLC test statistics. The horizontal dash line indicates there is no difference between the two statistics and the vertical dash line is the critical value for the F distribution at the nominated $5\%$ level.}
\label{fig:MC_QB}
\end{figure}

\begin{figure}[H]
 \begin{center}	
\includegraphics[width=17cm]{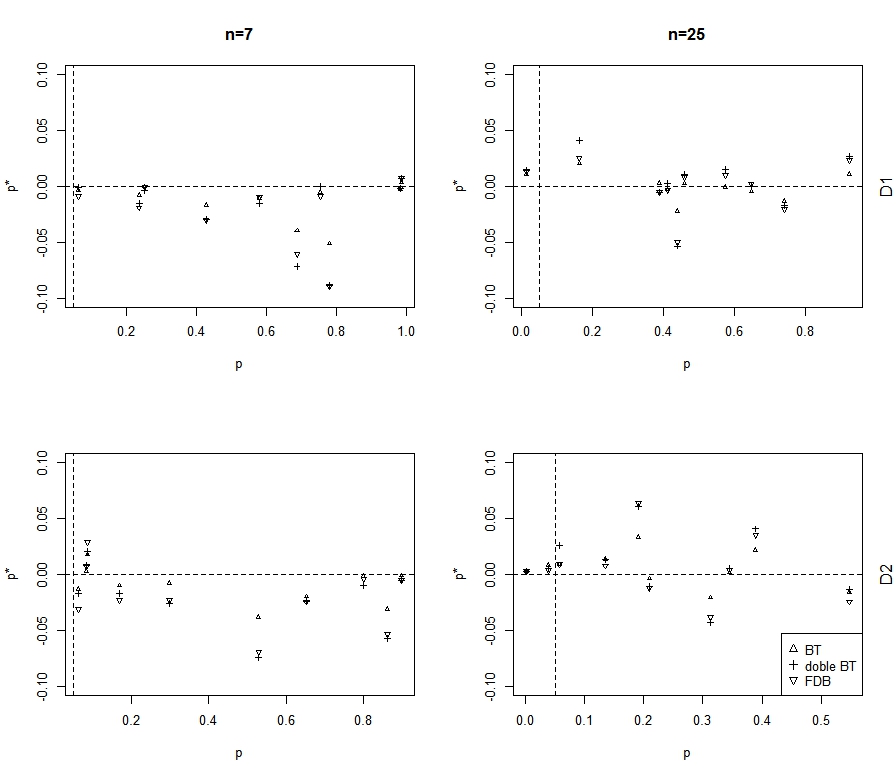}
 \end{center}
\caption{Plots of P values from various bootstrap methods against theirs FLC test counterpart for $10$ datasets generated using the Setting $2$ with Student's t distribution with 3 degrees of freedom, when the size of the cluster $m=10$. Bootstrap methods compared included the null residual bootstrap (BT), the double bootstrap with $999$ first level and $150$ second level bootstrap samples (double BT) and the fast double bootstrap (FDB). Each row represents one of two covariance matrices $\mathbf{D}_{1}=$~\usebox{\smlmatit} and $\mathbf{D}_{2}=$~\usebox{\smlmatiit}. The y-axis shows the differences between the P value from a given bootstrap method and its corresponding P values calculated from a FLC test. The horizontal dash line indicates there is no difference between the two P values and the vertical dash line is when P value is $0.05$.}
\label{fig:MC_p}
\end{figure}

A limited number of simulations were conducted to examine the variability of $\hat{Q}_{B}^{**}$ of the fast double bootstrap (Figure \ref{fig:MC_QB}) and the difference between the double bootstrap and the fast double bootstrap (Figure \ref{fig:MC_p}). We generated $10$ datasets using Setting $2$ and considered the errors were individually and independently distributed with Student's t distribution with 3 degrees of freedom. We chose two different numbers of clusters $(n=7\mbox{ or }n=25)$ and two covariance matrices for subset of the random effects: one for when the null hypothesis is true, i.e., $\mathbf{D}_{1}=$~\usebox{\smlmatit}; and another one for when the null hypothesis is not true, i.e., $\mathbf{D}_{2}=$~\usebox{\smlmatiit}. For each dataset, we conducted $10$ Monte-Carlo simulations using the fast double bootstrap, and these $1-\hat{p}^*$ quantile $\hat{Q}_{B}^{**}$ were subtracted off their corresponding FLC test statistic and plotted in Figure \ref{fig:MC_QB}.

Generally speaking, the variability of the second-level quantiles seemed to grow slightly with the underlying FLC test statistics. The increase in the number of cluster size (and the sample size) reduced the variation in $\hat{Q}_{B}^{**}$. Note that the scale of the y-axis is twice as much as of the x-axis. This choice is deliberate for showing the small variability in most of the cases. The worst case is a $0.2$ difference between the $\hat{Q}_{B}^{**}$ and the FLC test statistic when the FLC test statistic is around $1.9$ in the bottom right plot for $n=25$ and $\mathbf{D}_{2}$.

Figure \ref{fig:MC_p} simultaneously compare the outcomes of the three bootstrap methods to the FLC test , namely, the residual bootstrap, the double bootstrap and the fast double bootstrap. 
For each dataset, we simulated $B=999$ bootstrap samples for the residual bootstrap and the fast double bootstrap. For the double bootstrap, we considered $B_1=999$ and $B_2=150$ as the numbers of bootstrap samples simulated for the first-level and seconde-level resamplings respectively. Note that the P values for the fast double bootstrap were the averages of the P values calculated from the $10$ Monte-Carlo simulations of the fast double bootstrap.

The results showed that the P values from the residual bootstrap generally were the closest to the P values from the FLC test. The P values from the two bootstrap methods with secondary adjustment further diverged from the P values of the FLC tests. However these differences became negligible for the datasets with small difference between the residual bootstrap and the FLC test. In addition, there was no notable difference between the double bootstrap and the fast double bootstrap.

In summary,
\begin{itemize}
  \item the variability of $\hat{Q}_{B}^{**}$ increases as the FLC test statistics increases for the fast double bootstrap;
  \item the differences between all three bootstrap methods are minimal when the P value from the residual bootstrap is close to the P values from the FLC test; and
  \item despite the computational differences, there is no notable difference in P values between the double bootstrap and the fast double bootstrap.
 \end{itemize}

\bibliographystyle{chicago}
\bibliography{bibliography}

\end{document}